\documentclass[preprint,groupedaddress,prb,onecolumn,11pt]{revtex4}
\usepackage{amsmath}
\usepackage{graphicx}

\begin{document}

\title{Thermal conductivity of deformed carbon nanotubes}
\author{Wei-Rong Zhong$^{1}$, Mao-Ping Zhang$^{1}$, Dong-Qin Zheng$^{1}$,
and Bao-Quan Ai$^{2}$}
\email{wrzhong@hotmail.com, aibq@scnu.edu.cn}
\affiliation{$^{1}$\textit{Department of Physics and Siyuan Laboratory, College of
Science and Engineering, Jinan University, Guangzhou 510632, P. R. China}}
\affiliation{$^{2}$\textit{Laboratory of Quantum Information Technology, ICMP and SPTE,
South China Normal University, Guangzhou, 510006 P. R. China}}
\date{\today}

\begin{abstract}
We investigate the thermal conductivity of four types of deformed carbon
nanotubes by using the nonequilibrium molecular dynamics method. It is
reported that various deformations have different influence on the thermal
properties of carbon nanotubes. For the bending carbon nanotubes, the
thermal conductivity is independent on the bending angle. However, the
thermal conductivity increases lightly with XY-distortion and decreases
rapidly with Z-distortion. The thermal conductivity does not change with the
screw ratio before the breaking of carbon nanotubes but decreases sharply
after the critical screw ratio.
\end{abstract}

\pacs{65.80.-g Thermal properties of small particles, nanocrystals,
nanotubes, and other related systems}
\pacs{ 81.07.De Nanotubes }
\maketitle

Due to lots of potential applications of carbon nanotubes (CNTs), more and
more attentions have been given to the exceptional and unique electronic and
thermal properties of CNTs in the past few years \cite{Pop1}$^{,}$ \cite%
{Baughman}$^{,}$ \cite{Jorio}$^{,}$ \cite{XLi}. Berber et al \cite{Berber}
applied molecular dynamics (MD) simulations to study the thermal
conductivity of CNTs. They found that CNTs show an unusually high thermal
conductivity value of 6600W/mK at room temperature. Pop et al \cite{Pop2}
represented an experimental estimate that the thermal conductance is
approximately 2.4 nW/K, and the thermal conductivity is nearly 3500 W/mK at
room temperature for a SWNT of length 2.6 $\mu m$ and diameter 1.7 nm. The
result of MD calculation is about two times that of experimental estimate.

Generally, the differences between experiment and theory can be explained by
two aspects. Firstly, the thermal characteristics extracted in experiment
must rely on an assessment of the SWNT temperature, which is very difficult
to obtain experimentally in a quantitative manner \cite{Pok}$^{,}$ \cite{Xu}%
. Secondly, CNTs used in MD simulations are perfect structures, however,
perfect CNTs are hardly to obtain through experiment. The former will depend
on the improvement of the experimental conditions. The later, however,
should be attributed to the theoretical reasons.

On the other hand, as have been shown by many simulations and experiments,
perfect carbon nanotubes (CNTs) perform very good thermal conductivity.
However, due to the need of application, it is inevitable that the CNTs
would be bended, twisted, stretched and compressed when they are used.
Recently, a few studies have reported that imperfect structures, such as
defects, deformations and impurities, can affect the thermal conductivity of
nanotube significantly \cite{CTang}$^{,}$ \cite{GWu1}$^{,}$ \cite{Lu}$^{,}$
\cite{Pease}$^{,}$ \cite{WZhang}. However, no investigations address on the
quantitative relationship between the imperfect parameter and the thermal
conductivity of CNTs.

In this work we first study the thermal conductivity of four types of
deformed carbon nanotubes by using the nonequilibrium molecular dynamics
method. Four typical parameters, i.e., the bending angle $\theta $, the
screw ratio $\beta /Z_{0}$, the stretch factor $Z^{\prime }/Z_{0}$ and
ellipse ratio $e$, are defined to describe the deformed degrees of CNTs. By
calculating the thermal conductivity of CNTs in various deformed parameters,
we try to get the relationship between the thermal conductivity and the
deformed parameters, which will be of significance to understand the role of
the deformation in the thermal properties of CNTs.

Figure 1 depicts four typical deformed carbon nanotubes. Fig.1(a) shows a
normal carbon nanotube. In our simulations, the nanotubes are deformed from
two types (zigzag (7,0) and armchair (7,7)) carbon nanotubes with the normal
length $Z_{0}$=18 nm. Fig.1(b) illustrates a bending carbon nanotube with a
bending angle $\theta $, here $\theta $ is the intersection angle between
the normal direction of the two end sections of the tube. If twisting the
nanotube around the Z-axial direction, one can obtain a twisted nanotube. As
shown in Fig.1(c), the screw distortion is denoted to the ratio of the
twisted angle ($\beta $) to the length of the nanotube ($Z_{0}$).
Z-distortion is the tension and compression along Z-axial direction. As
shown in Fig.1(d), the stretch factor is Z$%
{\acute{}}%
$/Z$_{0}$, where Z%
\'{}
and Z$_{0}$ are respectively the length of the distortion nanotube and the
normal nanotube. Normally, the width of X-axial direction of the carbon
nanotube equals to that of Y-axial direction. When applying a force on X- or
Y-direction, the nanotube will become a XY-distortion nanotube. As
illustrated in Fig.1(e), for XY-distortion nanotube, the ellipse ratio \cite%
{Harris}, also called the eccentricity of an ellipse, usually denoted by $e$%
, is the ratio of the distance between the two foci, to the length of the
one axis or $e=2f/2a=f/a=f/Y$.
\begin{figure}[htbp]
\begin{center}\includegraphics[width=8cm,height=6cm]{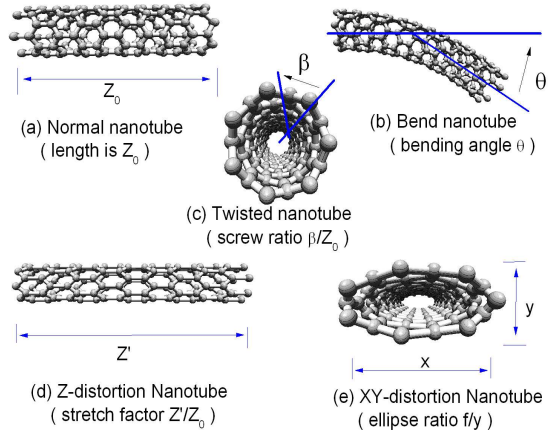}
  \end{center}
  \caption{
Normal and four types of deformed carbon nanotubes.}
   \label{}
\end{figure}

In our simulations, we have used classical molecular dynamics method based
on the Tersoff-Brenner potential \cite{Tersoff} of C-C bonding interactions.
The equations of motion for atoms in either the left or right Nos\'{e}%
-Hoover thermostat are \cite{Nose}$^{,}$ \cite{Hujin}$^{,}$ \cite{GWu2}%
\begin{equation}
\frac{d}{dt}p_{i}=F_{i}-\gamma p_{i};\frac{d}{dt}\gamma =\frac{1}{Q}\left[
\underset{i}{\sum }\frac{p_{i}^{2}}{2m_{i}}-3Nk_{B}T_{0}/2\right] ,
\end{equation}%
where $p_{i}$ is the momentum and $F_{i}$ is the force applied on the $i$-th
atom. $Q=3Nk_{B}T_{0}\tau ^{2}/2$, where $\tau $ is the relaxation time,
which is kept as $1ps$. $\gamma $ is the dynamic parameter of the
thermostat, $T(t)$,\ which is defined as $\frac{m_{i}}{3k_{B}}%
(vx(t)^{2}+vy(t)^{2}+vz(t)^{2})$, where $v(t)$ is the time-dependent
velocity, is the instant temperature of the heat baths at time $t$. $T_{0}$ (%
$T_{H}$ or $T_{C}$), the set temperature of the heat baths, are placed at
the two ends of a carbon nanotube and the temperature difference is denoted
as $\Delta T=T_{H}-T_{C}$. For the convenience of comparison, here we set $%
T_{H}=330K,T_{L}=290K$, which are near the room-temperature. In order to
keep the shape of the deformed nanotube in the simulations, we use fixed
boundary conditions in the two ends of the nanotube. The fixed region and
the heat baths occupy four layers and eight layers of atoms, respectively. $%
N $ is the number of the atoms in the heat baths, $k_{B}$ is the Boltzmann
constant and $m$ is the mass of the carbon atom.

These equations of motion are integrated by Verlet method \cite{Rafii}. The
time step is $0.50$ $fs$, and the simulation runs for $2\times 10^{7}$ time
steps giving a total molecular dynamics time of $10$ $ns$. Generally, $T(t)$
can stabilize around the set value $T_{0}$ after 1 ns. Time averaging of the
temperature and heat current is performed from 5 to 10 ns. The heat bath
acts on the particle with a force $-\gamma p_{i}$; thus the power of heat
bath is $-\gamma p_{i}^{2}/m$, which can also be regarded as the heat flux
from heat bath. The total heat flux injected from the heat bath to the
system can be obtained by \cite{GWu2}%
\begin{equation}
J=\underset{i}{\sum }\left[ -\gamma p_{i}^{2}/m_{i}\right] =-3\gamma
Nk_{B}T(t),
\end{equation}%
where the subscript $i$ runs over all the particles in the thermostat. The
final thermal conduction is calculated from the well-known Fourier's law $%
\kappa =Jl/(\Delta Tch)$, in which $l$, $c$, and $h$ (=0.34nm) are the
length, perimeter and thickness of the carbon nanotubes, respectively.

In order to investigate the chirality dependence of the thermal
conductivity, (7,0) and (7,7) deformed CNTs are studied simultaneously in
our simulations. There are 1197, 2072 atoms, respectively in the (7,0) and
(7,7) CNTs. The deformed parameters of the CNTs include the bending angle,
the screw ratio, the stretch factor and the ellipse ratio.

First, we calculate the thermal conductivity of the bending zigzag and
armchair CNTs. Figure 2 shows that the thermal conductivity alters slightly
with the bending angle not only for zigzag tubes but also for armchair
tubes. At the ranges of bending angle from 2 to 50 degrees, the values of
thermal conductivity change respectively from 1689 to 1734W/mK for zigzag
CNTs, and from 2263 to 2400W/mK for armchair CNTs. The fluctuation
respectively ranges less than 2.6\% and 5.7\%. The result shows that the
thermal conductivity is independent on the bending angle in the range of
error.
\begin{figure}[htbp]
\begin{center}\includegraphics[width=8cm,height=6cm]{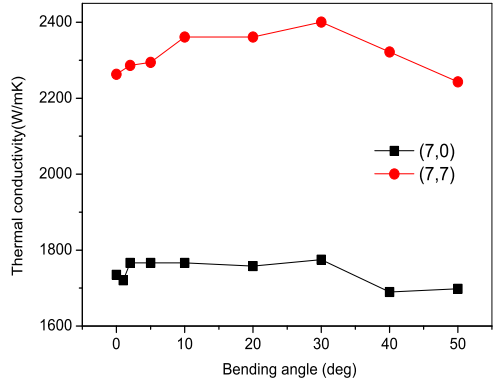}
  \end{center}
  \caption{Thermal conductivity of
(7,0) and (7,7) bending carbon nanotubes.}
   \label{}
\end{figure}

We have also investigated the twisted CNTs and found that the screw ratio
has no significant influence on the thermal conductivity at the small range
of its value. As shown in Fig. 3, for armchair tubes, the thermal
conductivity does not change with the screw ratio within the critical value $%
\beta /Z_{0}=3.0$ $deg/nm$. For zigzag tubes, the critical value is about $%
15.0\ deg/nm.$ Different behaviors between the zigzag and armchair graphene
(or nanotubes) have also been studied by the use of classical as well as
quantum molecular dynamics method \cite{Shen}$^{,}$ \cite{Gao}. In these
studies, it is suggested that the critical loading of the zigzag graphene is
larger than that of the armchair graphene. Our result exactly, at the view
of thermal transport, confirmed that the zigzag CNTs have better mechanical
properties than the armchair CNTs. Two states of the zigzag CNTs, state (a)
and state (b) in Fig. 3, which are the states before and after breaking
respectively, are plotted in Figs. 4(a) and (b). These states correspond to
the shapes of CNTs after 10 ns simulations. Similar three states of armchair
CNTs in Fig. 3 (points (c), (d) and (e)) are shown in Figs. 4(c), (d) and
(e) respectively.

Generally, when the CNTs is twisted (before break off), the chirality of
tubes changes and then the atom chains are not parallel to the axis, instead
they are in a helix. The vibration of the atoms in deformed nanotubes is apt
to have a long path, which causes the low thermal conductivity of deformed
nanotubes \cite{Ren}$^{,}$ \cite{WZhang}$^{,}$ \cite{Lu}. On the other hand,
the ballistic transport is the mainly thermal transport mode for the CNTs at
this temperature and length \cite{Mingo}$^{,}$ \cite{JWang}. Since the
length of CNTs is invariable when being twisted, the thermal conductivity
keeps on a certain value just before the CNTs' break off.

When the screw ratio goes over the critical value, the tubes are breaking,
as shown in Figs. 4(b), (d) and (e), an junction occurs in the middle of the
tubes, which will increases the resistance of the thermal transport greatly,
and then the thermal conductivity drops rapidly. 
\begin{figure}[htbp]
\begin{center}\includegraphics[width=8cm,height=6cm]{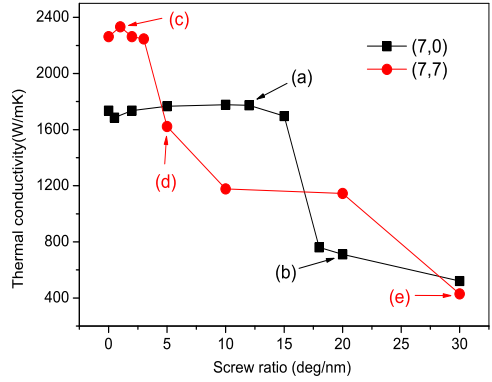}
  \end{center}
  \caption{Thermal conductivity of (7,0) and (7,7) twisted carbon
nanotubes. The structures of the CNTs correspond to the points (a), (b),
(c), (d) and (e) are displayed in Figs.4(a), (b), (c), (d) and (e)
respectively. }
   \label{}
\end{figure}

\begin{figure}[htbp]
\begin{center}\includegraphics[width=8cm,height=6cm]{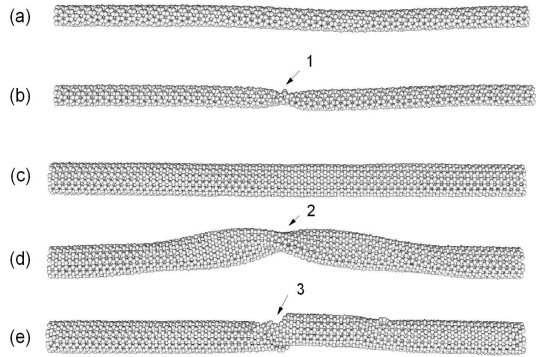}
  \end{center}
  \caption{The shape of CNTs after 5$\times 10^{7}$ steps of simulations. (7,0)
CNTs: screw ratio = 12 (a) and 20 (b) $deg/nm$, position 1 is the breaking
junction. (7,7) CNTs: screw ratio = 2.0 (c), 5.0 (d) and 30 (e) $deg/nm$,
position 2 and 3 are the breaking junctions.}
   \label{}
\end{figure}

The thermal conductivity of the Z-distortion as well as the XY-distortion
CNTs are also studied. Figure 4 illustrates that not only the tension but
also the compression can decreases the thermal conductivity of CNTs
strongly. When the stretch ratio $Z\prime /Z_{0}=0.9$, the thermal
conductivity decreases to 45\% and 56\% of the normal value for armchair and
zigzag CNTs, respectively. This phenomenon can be explained to negative
influence of the stretching on the full optimization of CNTs. For
XY-distortion CNTs, as displayed in Fig.5, the thermal conductivity
increases slightly when the ellipse ratio is lower or larger than 1. The
section area increases as the shape of the CNTs changes from circle to
ellipse, which helps the improvement of the thermal conductivity.
\begin{figure}[htbp]
\begin{center}\includegraphics[width=8cm,height=6cm]{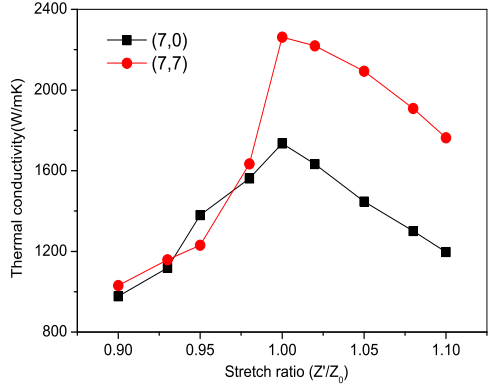}
  \end{center}
  \caption{Thermal conductivity of (7,0) and
(7,7) Z-distortion carbon nanotubes.}
   \label{}
\end{figure}

\begin{figure}[htbp]
\begin{center}\includegraphics[width=8cm,height=6cm]{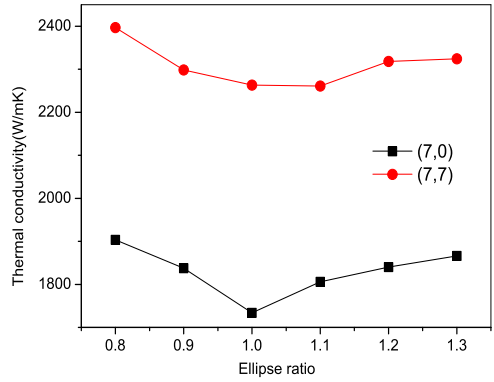}
  \end{center}
  \caption{Thermal conductivity
of (7,0) and (7,7) XY-distortion carbon nanotubes.}
   \label{}
\end{figure}

Here we have to point out that all deformed parameters set in our
calculations are small enough to guarantee against the occurrence of CNTs'
snap. For large deformed CNTs, the CNTs will go to a mechanical critical
state, and then it is difficult to describe the deformed degrees of CNTs
accurately. On the other hand, the temperature we use in the paper is the
temperature ($T_{MD}$) calculated from the classical MD, which can be
corrected by taking into account the quantum effects of phonon occupation,
using the expression $T_{MD}=(2T^{3}/T_{D}^{2})\int_{0}^{T_{D}/T}\left[
x^{2}/(e^{x}-1)\right] dx$, where $T$ and $T_{D}$ (322 K) are the corrected
temperature and Debye temperature, respectively \cite{Hujin}.

In summary, we have studied the thermal conductivities of four typical
deformed CNTs, using the nonequilibrium MD method based on the
Tersoff-Brenner potential. We observed the deformation dependence of the
thermal conductivity of CNTs. The thermal conductivity of the bending CNTs
does not change with the bending angle. Before the breaking of CNTs, the
thermal conductivity is independent on the screw ratio. When the twisting
goes over the mechanical criticality, the thermal conductivity decreases
rapidly. From the thermal properties of the twisted CNTs, it is suggested
that zigzag CNTs maybe have better mechanical property than armchair CNTs.
For the distortion in longitudinal and transverse direction, the thermal
conductivity increases lightly with XY-distortion and decreases rapidly with
Z-distortion.

\begin{acknowledgments}
We would like to thank Siyuan cluster for running part of our programs. This
work was supported in part by the National Natural Science Foundation of
China (Grant No.11004082) and the Natural Science Foundation of Guangdong
Province, China (Grant No.01005249).
\end{acknowledgments}

\end{document}